\documentclass{iopart}

\usepackage{mathbbol,amssymb,latexsym,amsthm}
\usepackage{iopams}
\usepackage{color,graphicx}
\usepackage{textcomp}

\def\a{\alpha}

\def\g{\gamma}

\def\e{\mathrm{e}}

\def\i{\mathrm{i}}

\def\D{\Delta}

\def\p{\partial}

\def\C{\mathbb C}

\def\R{\mathbb R}
\def\N{\mathbb N}

\newcommand{\I}{\Eins}
\newcommand{\ket}[1]{\left\vert #1\right\rangle}
\newcommand{\bra}[1]{\left\langle #1\right\vert}

\theoremstyle{plain}
\newtheorem{thm}{Theorem}[section]

\newtheorem{prop}[thm]{Proposition}
\theoremstyle{definition}

\theoremstyle{remark}

\begin{document}

\title{Transport of interface states in the Heisenberg chain}

\author{Tom Michoel$^1$, Bruno Nachtergaele$^2$ and Wolfgang Spitzer$^3$}

\address{$^1$ Institute for Theoretical Physics, Katholieke
  Universiteit Leuven, Celestijnenlaan 200D, B--3001 Leuven,
  Belgium.\footnote{Current address: Department of Molecular Genetics,
    Ghent University and Department of Plant Systems Biology, VIB,
    Technologiepark 927, B-9052 Gent, Belgium.\newline \copyright\
    2008 by the authors. This article may be reproduced in its
    entirety for non-commercial purposes.} }

\address{$^2$ Department of Mathematics, University of California,
  Davis, One Shields Avenue, Davis, CA 95616-8366, USA.}

\address{$^3$ Institut f\"ur Theoretische Physik, Universit\"at
  Erlangen-N\"urnberg, 91058 Erlangen, Germany.}

\eads{\mailto{tom.michoel@psb.ugent.be},
  \mailto{bxn@math.ucdavis.edu},
  \mailto{wolfgang.spitzer@physik.uni-erlangen.de}}
 
\date{October 5, 2008}

\begin{abstract} 
  We demonstrate the transport of interface states in the one-dimensional
  ferromagnetic Heisenberg model by a time dependent magnetic
  field.  Our analysis is based on the standard Adiabatic Theorem.
  This is supplemented by a numerical analysis via the recently
  developed time dependent DMRG method, where we calculate the
  adiabatic constant as a function of the strength of the magnetic
  field and the anisotropy of the interaction.
\end{abstract}

%\maketitle

%%%%%%%%%%%%%%%%%%%%%%%%%%%%%%%%%%%%%%%%%%%%%%%%%%%%%%%%%%%%%%%%%%%%%%%%%%%%%%%%%%%%%%%%%%%%%
\section{Introduction}
%%%%%%%%%%%%%%%%%%%%%%%%%%%%%%%%%%%%%%%%%%%%%%%%%%%%%%%%%%%%%%%%%%%%%%%%%%%%%%%%%%%%%%%%%%%%%

The problem of calculating the dynamics of quantum spin chains has
acquired new importance due to potential new applications in
spintronics~\cite{Allwood,beach05}, quantum information, computation
and control theory and the now realistic possibility of doing
experiments on systems that are accurately described by a
one-dimensional array of spins~\cite{Ono,hirjibehedin06}. The
development of time-dependent versions of the successful Density
Matrix Renormalization Group (DMRG) method of White
\cite{white1993,peschel99} by several authors \cite{vidal2004, Daley,
  white2004,gobert05} therefore comes at a fortuitous moment.  All
numerical schemes for a quantum many-body problem such as a spin chain
necessarily involve a drastic reduction of the high-dimensional
Hilbert space to a suitable subspace of relatively modest dimension.
The DMRG method does this by approximating the desired states
(typically the ground state and low-lying excitations above it) by
finitely correlated states \cite{FNW1}, also known as generalized
valence bond states or matrix product states. The time-dependent DMRG
method extends this idea by allowing the subspace used in the
approximation scheme to depend on time.

Here we are interested in the dynamics of the Heisenberg XXZ chain
with kink boundary conditions \cite{PS} and a time-dependent external
field. The ground states of this model are the so-called kink states
which describe an interface between domains of opposite magnetization.
Their degeneracy grows with the length of the chain \cite{ASW,GW}, and
is labelled by the position of the kink, or equivalently, the total
magnetization which is a conserved quantity. A transverse external
field localized at one site can be used to select a unique ground
state \cite{CNS}, i.e., such a perturbation pins the kink at a
specific site. These interface structures also appear in certain
asymmetric simple exclusion models of stochastic particle dynamics
\cite{Gwa1992,Derrida1993,Sandow1994} and it is a natural question to
study their non-equilibrium properties.  In this paper we study the
quantum dynamics of the kink under the influence of a time-dependent
magnetic field.  Typically we use a field supported at one or two
sites and moving at a constant speed. For small velocities, we know by
the adiabatic theorem that the time-evolution of a ground state will
follow the ground state of the time-dependent Hamiltonian. The
standard heuristic scale to identify the adiabatic regime is the smallness
of the ratio of the time-derivative of the Hamiltonian over the
spectral gap squared. We will verify the usefulness of this criterion
by calculating the time evolution of kink states in the XXZ chain
under the influence of a time-dependent magnetic field. Needless to
say, the dynamics of kink states, which can be regarded as discrete
analogues of solitons, is of great interest in its own
right. Moreover, they provide a simple dynamical quantum model of
sharp magnetic domain walls.

In order to numerically study the quality of the adiabatic
approximation and its domain of validity, it is necessary to express
the evolved state in the time-dependent eigenbasis of the Hamiltonian.
Therefore we construct the DMRG basis at time $t$ starting from the
ground state(s) and low-lying excited states of the Hamiltonian of the
system at time $t$. This is in contrast with the original method,
where the time-dependent DMRG basis is obtained from the evolved state
at time $t$. Our method can be of more general interest, since the
calculation of the DMRG basis is a separate calculation about which we
can be confident to have good control, regardless of the potential
difficulties in reliably calculating the time evolution.

The paper is organized as follows. In Section \ref{sec:model}, we
define the model and study its basic properties. The adiabatic
approximation is studied in Section \ref{Sect:adiabatic}. In Section
\ref{sec:fast-change-magnetic}, we give a brief discussion of the
issues that arise with rapidly changing magnetic fields. Details of
the DMRG algorithm are given in \ref{sec:algorithm}.

%%%%%%%%%%%%%%%%%%%%%%%%%%%%%%%%%%%%%%%%%%%%%%%%%%%%%%%%%%%%%%%%%%%%%%%%%%%%%%%%%%%%%%%
\section{The model}\label{sec:model}
%%%%%%%%%%%%%%%%%%%%%%%%%%%%%%%%%%%%%%%%%%%%%%%%%%%%%%%%%%%%%%%%%%%%%%%%%%%%%%%%%%%%%%%

The spin-$\frac{1}{2}$ ferromagnetic XXZ Heisenberg model on the chain
$[1,L]$ with interface boundary conditions is defined by the
Hamiltonian $H_0 = \sum_{x=1}^{L-1} h_{x,x+1}$, with nearest-neighbor
interaction
\begin{equation}\fl
  h_{x,x+1} = -\D^{-1}(S_x^1S_{x+1}^1 + S_x^2S_{x+1}^2) - S_x^3S_{x+1}^3 
  + \case12 \sqrt{1-\D^{-2}} (S_x^3 - S_{x+1}^3) + \case{1}{4}\I\,,
\end{equation}
where $\D>1$ is the anisotropy parameter and the matrices
$(S_x^1,S_x^2,S_x^3)$ are the usual spin-$\frac{1}{2}$ matrices;
$h_{x,x+1}$ is a projection and $H_0$ has ground state energy zero. We
have set $\hbar=1$ which should be kept in mind when we define
adiabatic regimes in terms of the magnitude of the magnetic field and
the velocity.

The Hamiltonian $H_0$ acts on the Hilbert space ${\mathcal H}_L =
\C^{2^L}$. It has a large (quantum) symmetry with a similar multiplet
structure as the $SU(2)$-symmetric, isotropic model where $\D=1$.  In
the limit $\D\to\infty$ we obtain the Ising Model, $H^{\mathrm{Ising}}
=H_0(\D=\infty)$. The ground states and excited states (for finite $L$
and in the infinite volume limit $L\to\infty$) have been analyzed
completely in recent years~\cite{KN1,KN3,NS,NSS3,NSS4}. A convenient
basis of ground states are the so-called grand canonical ground
states, $\phi_c$. They are product states and depend on a complex
parameter $c\in\C\setminus\{0\}$:
\begin{equation} \label{prod}
  \phi_c =\bigotimes_{x=1}^L\left[ (1+|c|^2q^{2x})^{-1/2}
    \left(\begin{array}{c}1\\cq^{x}\end{array}\right) \right]\,.
\end{equation}
The state $\phi_c$ describes an interface state that is exponentially
localized at $x_0=-\ln{|c|}/\ln{q}$ with $q=\D-\sqrt{\D^2-1}\in(0,1)$.
The width of the interface depends on $\D$ and becomes sharp (i.e.,
the transition from up to down spin occurs across one bond) in the
Ising limit $\D\to\infty$ and flat as $\D\downarrow1$. There are no
interface states in the isotropic Heisenberg model.

We perturb the Hamiltonian by a magnetic field ${\mathbf
B}:[1,L]\rightarrow\R^3$ satisfying the locality condition:
\begin{description}
\item[(A1)] The support of ${\mathbf B}$ %, $\mathrm{supp}({\mathbf B})$, 
is finite, non-empty, and independent of $L$.
%\item[(A2)] The field is of the form ${\mathbf B}(x) =
%  (B^1(x),0,B^3(x))$, where $B^1$ is a non-zero, non-positive function.
\end{description}
Then we define the Hamiltonian
%By applying a global rotation around the $3$-axis, fields of a
%slightly more general form can be considered as well.  Then we define
\begin{equation}\label{H_V} 
  H_V = H_0 + V\quad\mathrm{with}\quad V = \sum_{x} {\mathbf B}(x) \cdot
  {\mathbf S}_x\,.
\end{equation}
We refrain from considering the infinite chain limit in detail. Condition 
{\bf (A1)} ensures that $H_V$ is still a well-defined semi-bounded operator 
in the GNS representation of the \emph{unperturbed} Hamiltonian. 
%The second assumption makes sure that
%the ground state of $H_V$ is an interface state, in the sense that to
%the left/right of the support of $V$, the profile is asymptotically
%equal to $-\frac12$, respectively $\frac12$. If ${\mathbf
%  B}(x)=(0,0,B(x))$ with $B(x)\ge0$, the ground state is the all
%spin-up state which is then part of the continuous spectrum of $H_V$,
%see the discussion in~\cite{CNS}.

The case of a magnetic field located at a single site $y$ has been
analyzed in~\cite{CNS}. Let $B_1+\i B_2\not=0, c_y=-(\|{\mathbf B}\| + 
B_3)(B_1+\i B_2)^{-1} q^{-y}$, and let $\phi_{c_y}$ be a state of the form
(\ref{prod}). This state $\phi_{c_y}$ has energy
$-\frac{1}{2}\|{\mathbf B}\|$, and since the ground state energy can
be at most shifted by this amount, we know that it is a ground state.
It is also the unique ground state which moreover has a uniform gap as
$L\to\infty$. If $\mathbf B = (B,0,0)$ with $B\not=0$, then $\phi_{c_y}$
is a kink localized at site $y$. If ${\mathbf B}(x)=(0,0,B(x))$ with $B(x)
\le0$, then the ground state is the all spin-up state which, as $L\to\infty$ 
is however not in the GNS kink Hilbert space; see the discussion 
in~\cite{CNS}. %In the present situation we generalize this to extended
%magnetic fields.

In the Ising limit $\Delta\to\infty$, ground states may be largely degenerate 
as for $V=0$. For example, consider the magnetic field ${\mathbf B}(1) = 
(1,0,0), {\mathbf B}(2) = (0,0,1)$, and zero otherwise. Then $H_V$ has 
$L-1$ ``kink" ground states of the form $\psi_{[1,2]}\otimes\psi_{[3,L]}$. Here,
$\psi_{[1,2]}$ is the ground state of the two-site Hamiltonian which has the 
form  $\psi_{[1,2]} = (0,c,0,-\sqrt{1-c^2})$ for some $c>0$. Note that 
$\langle\psi_{[1,2]}|S_2^3\psi_{[1,2]}\rangle = -1/2$. Thus we can choose 
$\psi_{[3,L]}$ to be an (Ising) kink state on the remaining sites of which there
are $L-1$.

Thus we consider fields ${\mathbf B}$ such that for all $x$ with 
${\mathbf B}(x)\not=0$, ${\mathbf B}(x)$ has a non-vanishing component in the 
$xy$-plane. This means that, in addition to \textbf{(A1)}, we assume the 
following:
\begin{description}
\item[(A2)] On the support of $V$, ${\mathbf B}(x)\not=(0,0,\pm 
\|{\mathbf B}(x)\|)$.
\end{description}
%Condition \textbf{(A2)} is certainly not optimal but satisfied by our examples
%that we consider later. 

\begin{thm} \label{thm:low energy kink} Let ${\mathbf B}$ satisfy the
  conditions \textbf{(A1)} and \textbf{(A2)}. Then, there exists a
  finite $\D_0=\D_0({\mathbf B})$ so that for all $\D>\D_0$,
\begin{enumerate}
\item $H_V$ defined in (\ref{H_V}) has a unique ground state;
\item $H_V$ has a positive gap above the ground state uniformly in
  $L$. I.e., there exists a $\gamma(\Delta)>0$, independent of $L$,
  such that for all states $\psi$ orthogonal to the ground state we
  have $\langle\psi|(H_V-E_0)|\psi\rangle \ge \g \,\|\psi\|^2$.
\end{enumerate}
\end{thm}

The conclusions of the theorem can be expected to be valid for other situations
than the ones covered by \textbf{(A1)} and \textbf{(A2)}. In the cases where we
have performed numerical calculations, a positive gap  
appeared for all $\D>1$. Here we only
show the result for sufficiently large $\Delta$ by applying the theory
of \emph{relatively bounded} perturbations, see e.g.~\cite{Thi3}. Let us
recall that an operator $B$ is relatively bounded with respect to the
operator $A$ if there exists a (finite) constant $M$ so that for all
$\psi$ in the domain of $A$, $\|B\psi\| \le
M\big(\|A\psi\|+\|\psi\|\big)$. Then, all eigenvalues and
eigenprojections in the discrete spectrum (i.e., isolated and finite
multiplicity) of $A(\a) = A + \a B$ are analytic for
$\a\in(-\a_0,\a_0)$ for some $\a_0>0$ (cf.~\cite[3.5.14]{Thi3}).

\begin{proof}

  Let $H_V^{\mathrm{Ising}} = H_V(\D=\infty)$ be the Ising kink
  Hamiltonian with magnetic field ${\mathbf B}$. By using a rotation around 
  the $z$-axis we can assume that ${\mathbf B}(x) = (B^1(x),0,B^3(x))$ with
  $B^1(x) < 0$ for all $x$ in the support $[a,b]$ of ${\mathbf B}$. Then the 
  Hamiltonian $H_{V,[a,b]}^{\mathrm{Ising}}$ defined on the support of 
  ${\mathbf B}$ has a unique ground state $\psi_{[a,b]}$ by the usual 
  Perron-Frobenius argument. I.e., there is an $n\in\N$ such that 
  $\big(H_{V,[a,b]}^{\mathrm{Ising}}\big)^n$ 
  has strictly negative off-diagonal entries. We claim now that $\psi_{[1,L]} 
  = \bigotimes_{x=1}^{a-1} {0\choose1}  \otimes \psi_{[a,b]} \otimes 
  \bigotimes_{x=b+1}^{L}{1\choose0} $ is the ground state of $H_V^{\mathrm{Ising}}$
  on the full chain $[1,L]$. By adding the other (non-negative) Ising
  Hamiltonians to the left and right of the support of ${\mathbf B}$ the
  ground state energy could in principle increase. But $\psi_{[1,L]}$ is also 
  an eigenvector of $H_V^{\mathrm{Ising}}$ with the same eigenvalue as
  $\psi_{[a,b]}$ since the additional terms vanish on $\psi_{[1,L]}$. Thus 
  $\psi_{[1,L]}$ is the ground state of $H_V^{\mathrm{Ising}}$.
  
  Next, we consider the Heisenberg Hamiltonian $H_V(\D) =
  H_V^{\mathrm{Ising}} + \D^{-1} K + (1-\sqrt{1-\D^{-2}})P$. $P$ only
  involves the boundary spins and is therefore uniformly bounded. $K$
  contains the $XY$ terms in the Hamiltonian and the norm of this term
  is of order $L$. In particular, it is not uniformly bounded but one
  can show that $K$ is relatively bounded with respect to
  $H_V^{\mathrm{Ising}}$. This was done in \cite{Mulherkar2007} for
  the Hamiltonian without $B$-field. The relative bound is uniform in
  $L$.  By our assumptions, the $B$-field defines a bounded operator
  and hence $K$ is also relatively bounded with respect to
  $H_V^{\mathrm{Ising}}$.  It follows that for $\D^{-1}$ in a
  non-empty interval $[0,\D_0^{-1}]$, the spectral projection for an
  interval of energies containing the unperturbed ground state energy
  is an analytic function of $\D^{-1}$. This implies the statement of
  the theorem.
\end{proof}

Our prime example are magnetic fields located on two adjacent sites,
$x_0=L/2$ and $x_0+1$. More precisely, let $v,\,B>0$ and define
\begin{eqnarray*}
  B(x,t) &= - B \cdot f(x-vt)\,,\\
  f(x) &= \cases{1-|x-x_0| & for  $x_0-1\le x\le x_0+1$\\
    0&otherwise\\}\,.
\end{eqnarray*}
The perturbation 
\begin{equation}\label{eq:V}
  V(t) = \sum_{x=1}^L B(x,t) S_x^1 = -(1-vt)\,B\,S_{x_0}^1 
  - vt\,B\,S_{x_0+1}^1 
\end{equation}
for $0\le t\le\tau=1/v$ satisfies the conditions \textbf{(A1)} and
\textbf{(A2)}. Hence, we know from Theorem \ref{thm:low energy kink}
that the Hamiltonian,
\begin{equation}\label{H(t)}
    H(t) = H_0 + V(t)
\end{equation}
has a unique ground state, $\varphi(t)$, with a positive gap. The
ground state energy $E_0(t,B)$ is analytic on $(0,v^{-1})$, also as
$L\to\infty$. It is well-known \cite[3.5.23]{Thi3} that the ground state energy $E_0(t,B)$ is
concave in $t$. By symmetry we also have $E_0(t,B)=E_0(v^{-1}-t,B)$
for $0\le t\le 1/(2v)$.  This implies $E_0(1/2v,B) = \max_t E_0(t,B)$.
We can get more information on $E_0(t,B)$ from the low energy spectrum
of the Hamiltonian with the magnetic field at a single site. For
simplicity we put $v=1$.

\begin{prop} 
  For $t>0$ let $g(t)$ be the gap above the ground state $\phi_{c_y}$
  of $H_y(t) = H_0 - tB S_y^1$. Then $g(t)$ is increasing in $t$
  (strictly speaking, $g(t)$ also depends on $L$ which we will tacitly
  ignore). Further, on $[0,1/2]$ we have for the Hamiltonian $H(t)$ 
  defined through (\ref{eq:V}) and (\ref{H(t)}),
  \begin{equation*}
    H(t) \ge \frac12 \,g(2t) \,
    \big(1-|\langle\phi_{c_y}|\phi_{c_{y+1}}\rangle|\big)\,.  
  \end{equation*}
\end{prop}

\begin{proof} 
  We may assume that $B>0$; otherwise apply a rotation.  That $g(t)$
  is increasing follows from the variational principle. To this end,
  let $0<t<t'$. Let $\psi$ be orthonormal to $\phi_{c_y}$ such that
  $\bra{\psi}H_y(t') + \frac12 t'B\ket{\psi} = g(t')$.  Then,
  \begin{equation*}
    \bra{\psi} H_y(t) + \frac12 tB \ket{\psi} \\= g(t') + (t-t') B
    \bra{\psi} \frac12 - S_y^1 \ket{\psi} \,\le\,g(t') \,.
  \end{equation*}
  Thus, $g(t) \le g(t') $.  The estimate on the ground state energy of
  $H(t)$ is based on Kitaev's Lemma \cite[Lemma 14.4]{Kitaev}.  In
  general, let $A_1$ and $A_2$ be two operators on some Hilbert space
  $\mathcal H$. By $G_{1,2}$ and $\Pi_{1,2}$ we denote the space of
  ground states and the orthogonal projections onto $G_{1,2}$,
  respectively.  We assume that $G_1\cap G_2 =\{0\}$, and that for
  some $\delta>0$, $A_{1,2} \ge \delta (1-\Pi_{1,2})$. Then,
  \begin{equation*}
    A_1 + A_2 \ge \delta \Big( 1 - \sup_{\psi_{1,2}\in G_{1,2}}
    \big|\langle\psi_1|\psi_2 \rangle\big|\Big) \,. 
  \end{equation*}
  For our proof it suffices to apply this inequality with
  \begin{eqnarray*} 
    A_1 &= \case12 H_0 - (1-t)B (S_y^1 - \case12 \I) 
    = \case12 H_y(2(1-t)) + \case12 (1-t)B \I\ge 0\,, \\
    A_2 &= \case12 H_0 - tB (S_{y+1}^1-\case12 \I) = 
    \case12 H_{y+1}(2t) + \case12 {tB} \I\ge 0\,.
  \end{eqnarray*} 
\end{proof}

We supplement this analysis of the ground state properties of the
family of Hamiltonians (\ref{H(t)}) with numerical DMRG calculations
for a chain of length $L=20$ and an anisotropy of $\Delta=2$. The
chain length is chosen such that for given $\Delta$, $L$ is much
larger than the width of the interface, but otherwise there are no
qualitative differences for different combinations of $\Delta$ and
$L$.  We distinguish three ranges of $B$-values with qualitatively
different ground state properties.

\begin{figure}
  \centering
  \includegraphics[width=0.5\linewidth]{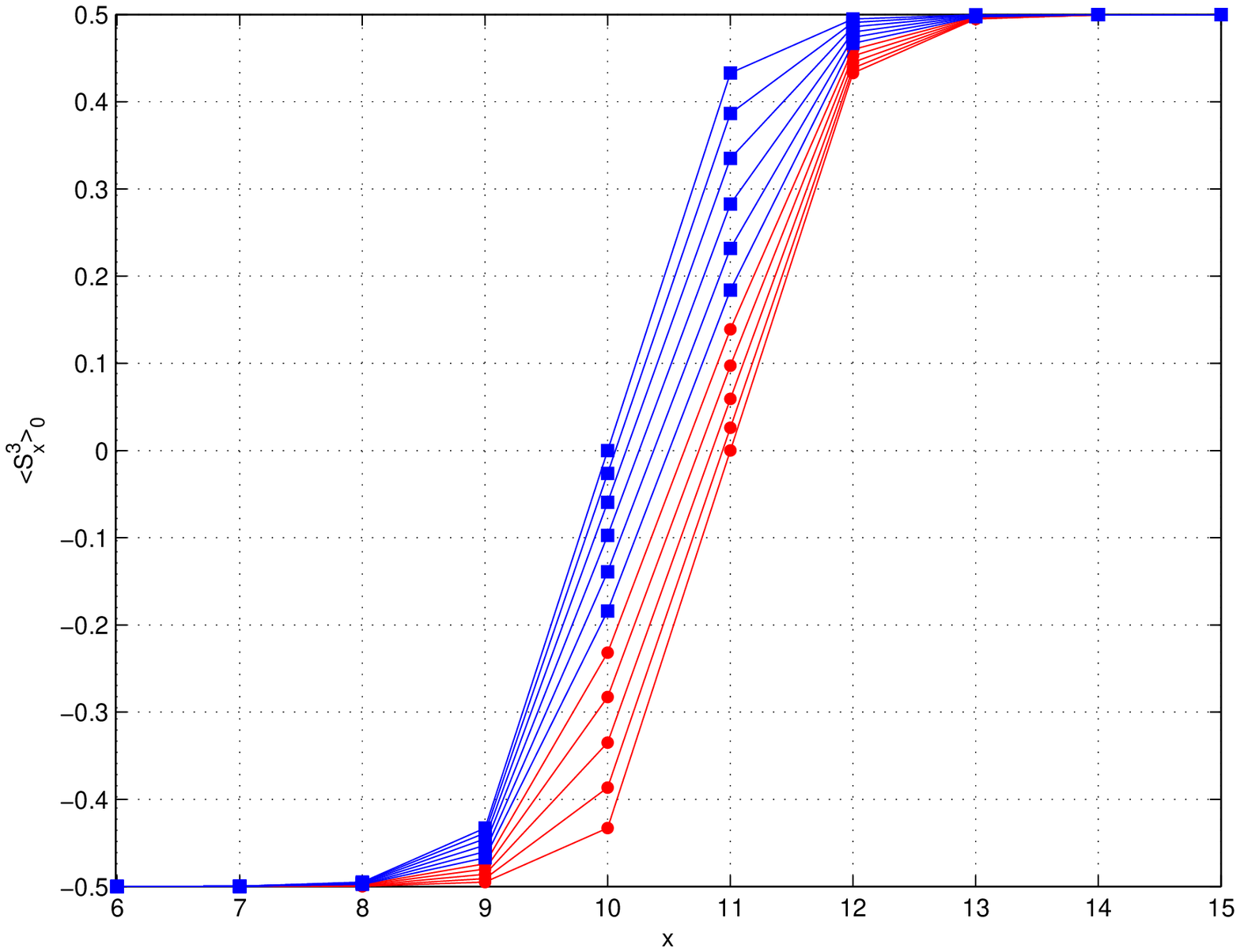}%
  ~\includegraphics[width=0.5\linewidth]{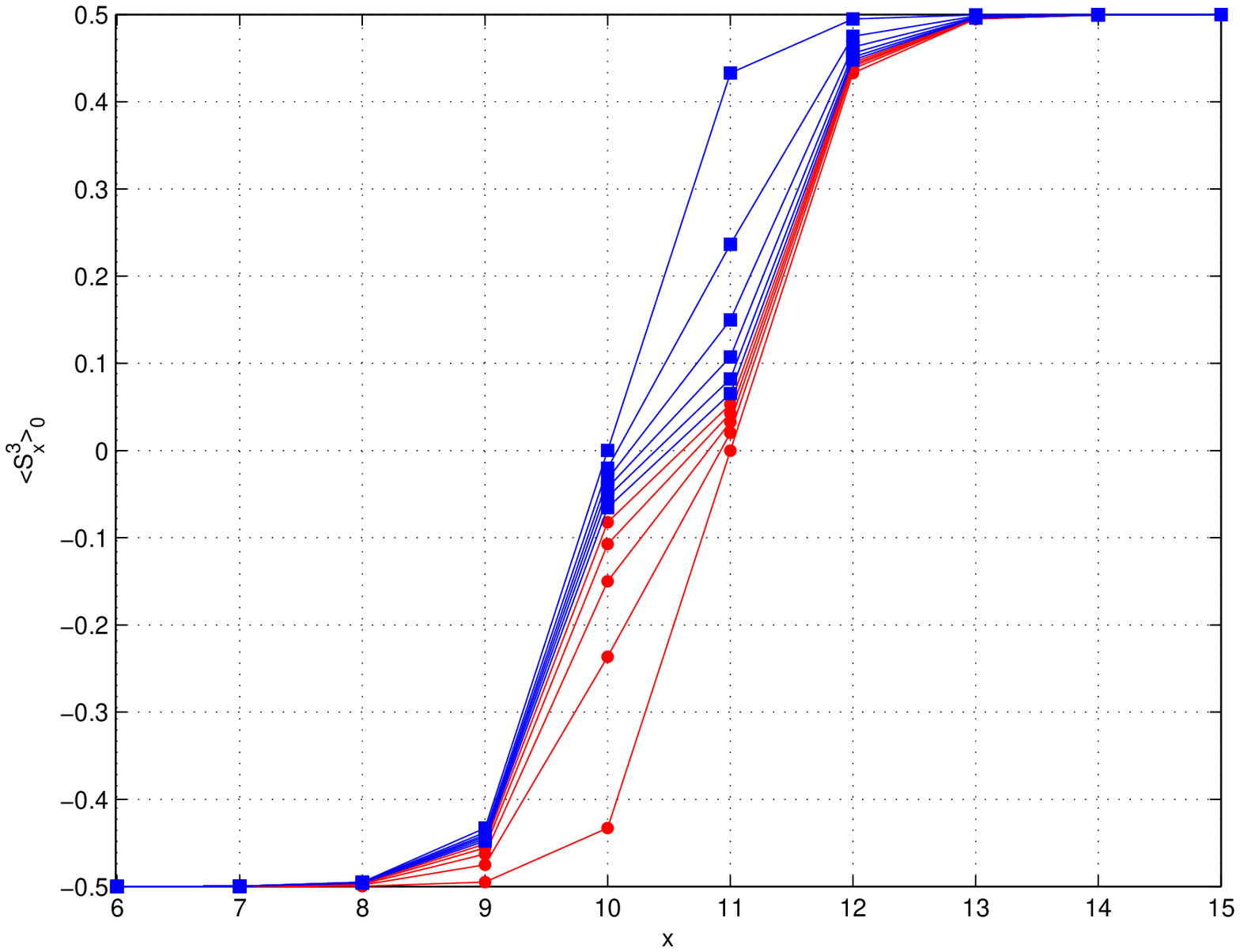}
  \caption{$S^3$-magnetization profiles of the ground states for
    $B=0.5$ (left) and $B=5$ (right), for $0\leq t\leq \tau/2$
    ({\color{blue}\tiny\fullsquare}) and $\tau/2\leq t\leq \tau$
    ({\color{red}\tiny\fullcircle}) ($\tau=200$, $200$ time steps,
    profiles plotted for every $20$ time steps, only central sites
    around the interface are shown).}
  \label{fig:S3_gs}
\end{figure}

\begin{figure}
  \centering
  \includegraphics[width=0.33\linewidth]{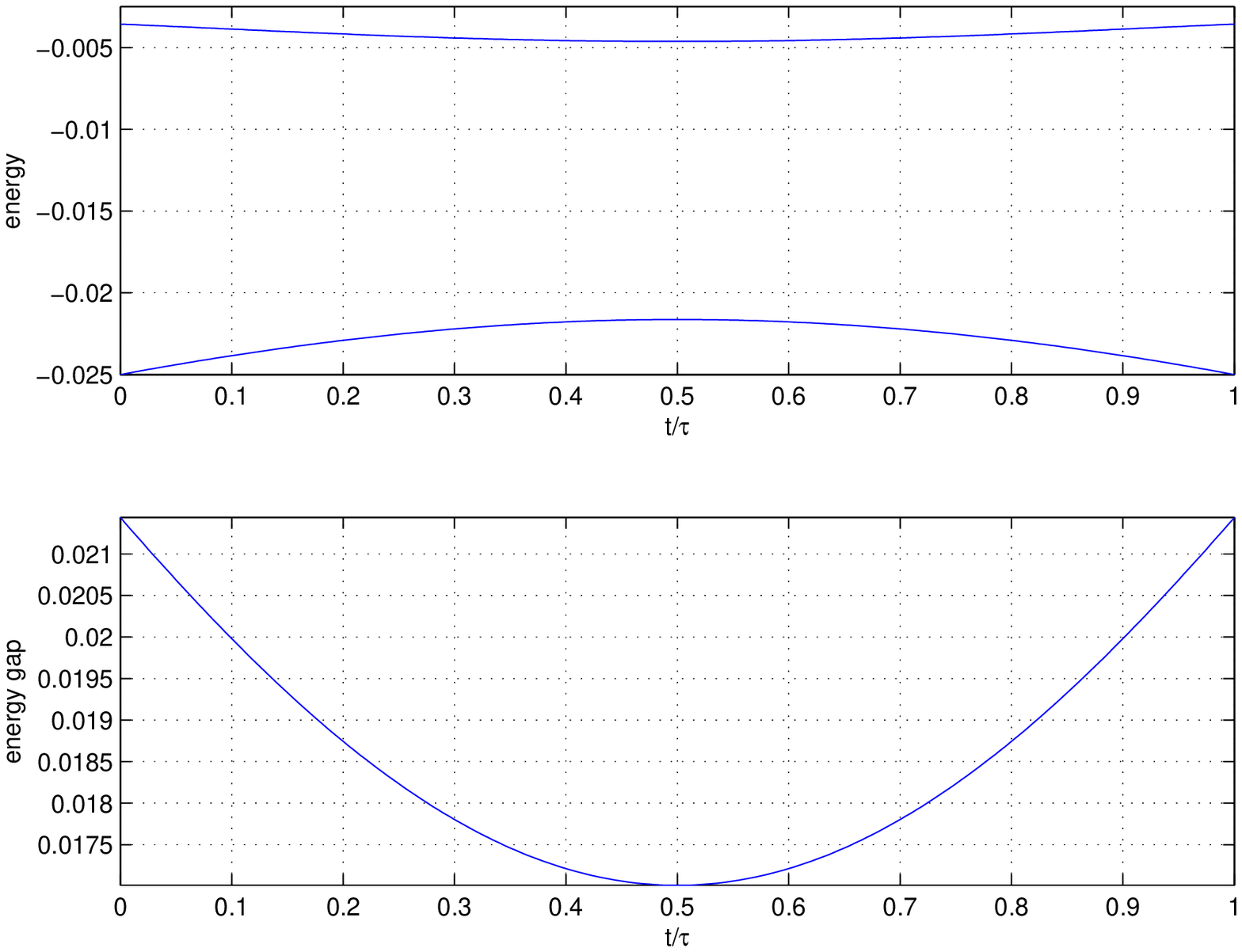}%
  ~\includegraphics[width=0.33\linewidth]{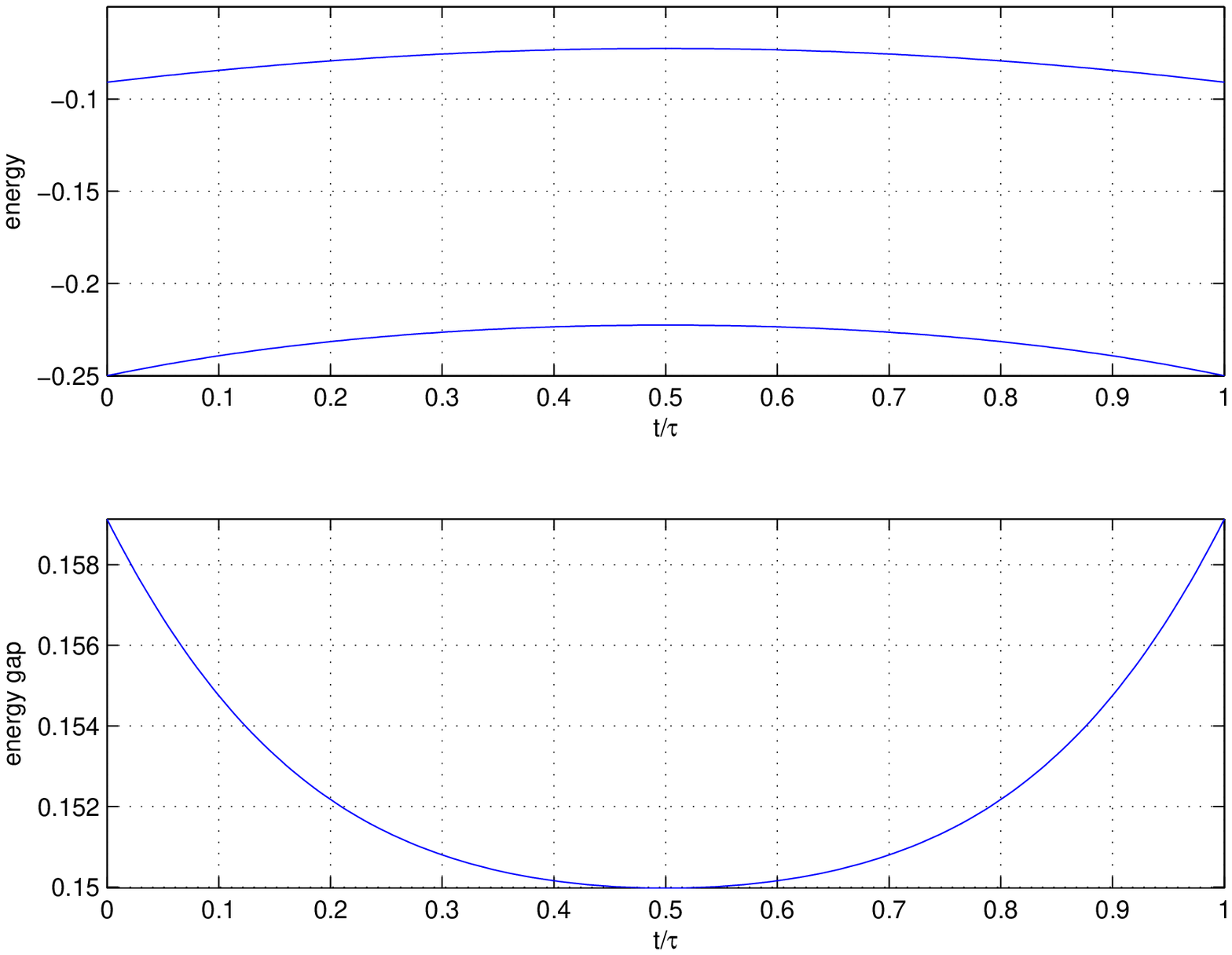}%
  ~\includegraphics[width=0.33\linewidth]{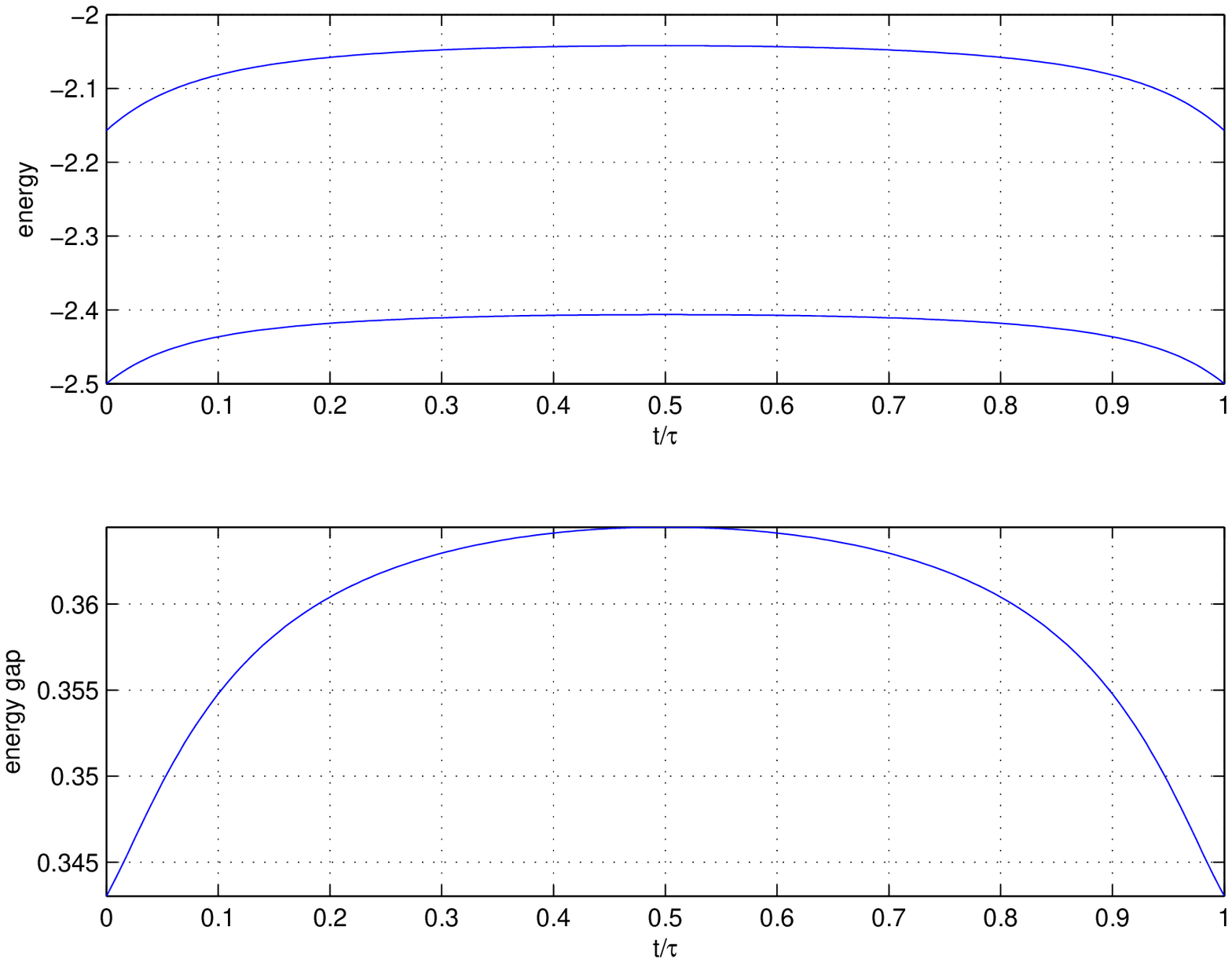}
  \caption{Energy of the ground states and first excited states (top)
    and energy gap (bottom) for $B=0.05$, $B=0.5$ and $B=5$ (left to
    right).}
  \label{fig:energy}
\end{figure}

A typical intermediate value of $B$ is $0.5$.  The $S^3$-magnetization
profile of the ground states $\varphi(t)$ of $H(t)$ nicely
interpolates between the interface product states
$\varphi(0)=\phi_{c_{x_0}}$ and $\varphi(1)=\phi_{c_{x_0+1}}$ (see
eq.~(\ref{prod})) with $x_0=L/2$ (Figure \ref{fig:S3_gs}).  The first
excited states are still well localized around the interface
positions. Both the ground state and first excited state energies are
concave as a function of $t$ (Figure \ref{fig:energy}).  The ground
state energy is minimal and equal to the known value $-B/2$ at the
start ($t=0$) and end ($t=\tau$) points.  The gap is always strictly
positive, but minimal at half period ($t=\tau/2$).  The transition
from $\varphi(0)$ to $\varphi(1)$ is not homogeneous, in the sense
that the right or top half of the interface moves first and the left
or lower half moves later. This effect becomes more pronounced as $B$
increases (compare Figure \ref{fig:S3_gs} left and right) and can be
understood as follows. In the first phase after switching on a
magnetic field at $x_0+1$, $S_{x_0+1}^1$ has the effect of rotating
the up vector at $x_0+1$ into the $(1,3)$-plane. But here nothing
happens at site $x_0$ because the initial state is an eigenstate of
$S_{x_0}^1$. The same thing happens at the end of the cycle when the
vector at $x_0+1$ is almost an eigenvector of $S^1_{x_0+1}$.  Site
$x_0$ is affected only after some time as the perturbation at $x_0+1$
is communicated to $x_0$ in second order.

For very small magnetic fields, the ground state energy is still a
concave function of $t$, but the first excited state energy is now
convex (Figure \ref{fig:energy}). The energy gap is still minimal at
half period. Furthermore the energy gap for a 10 times smaller
magnetic field ($0.05$ vs. $0.5$) is approximately 10 times smaller,
confirming the theoretical result \cite{CNS} (for a single site
impurity) that the gap scales linearly with $B$ for small $B$.  For
magnetic fields much larger than the intermediate value $B\approx
0.5$, the phenomenon that the interface moves in separate steps
becomes much more pronounced (Figure \ref{fig:S3_gs}). Like for
intermediate $B$-values, the energies of the ground and first excited
states are concave functions of $t$, but the minimal gap now occurs at
the start and final times (Figure \ref{fig:energy}).

%%%%%%%%%%%%%%%%%%%%%%%%%%%%%%%%%%%%%%%%%%%%%%%%%%%%%%%%%%%%%%%%%%%%%%%%%%%%%%%%%%%%%%%%%%%%%%%
\section{Adiabatic transport}\label{Sect:adiabatic}
%%%%%%%%%%%%%%%%%%%%%%%%%%%%%%%%%%%%%%%%%%%%%%%%%%%%%%%%%%%%%%%%%%%%%%%%%%%%%%%%%%%%%%%%%%%%%%%

Let us recall the standard adiabatic theorem which applies also to the
case of an infinite chain. Let $\tau>0$ and $H(t)$, $0\le t \le \tau$,
be a family of self-adjoint operators with common dense domain, and
let $\psi(t)$ be the solution to
\begin{equation}\label{SE} 
  \i \frac{\p}{\p t} \psi(t) = \tau H(t) \psi(t)
\end{equation}
with initial condition $\psi(0)$. By $P(t)$ we denote the spectral
projection onto the ground states of $H(t)$. We assume that $P(t)$ is
piece-wise, twice continuously differentiable, finite dimensional, and
uniformly (in $t$) separated from the rest of the spectrum of $H(t)$
by a gap $\gamma(t)$.

\begin{thm}[Adiabatic Theorem, Kato] \label{thm:adiab} Under
  the above conditions on $H(t)$ and $\psi(t)$, there is an
  eigenvector $\varphi(t)$ of $H(t)$ with $\varphi(0)=\psi(0)$ and a
  constant $C$ such that,
  \begin{equation}\label{equ:adiab}
    \sup_{0\le t\le\tau^{-1}} \|\psi(t) - \varphi(t)\| \le C \tau^{-1}\,.
  \end{equation}
\end{thm}

\begin{proof} 
  See~\cite{Kato} or \cite[3.3.11]{Thi3}.
\end{proof}

\begin{figure}
  \centering
  \includegraphics[width=0.5\linewidth]{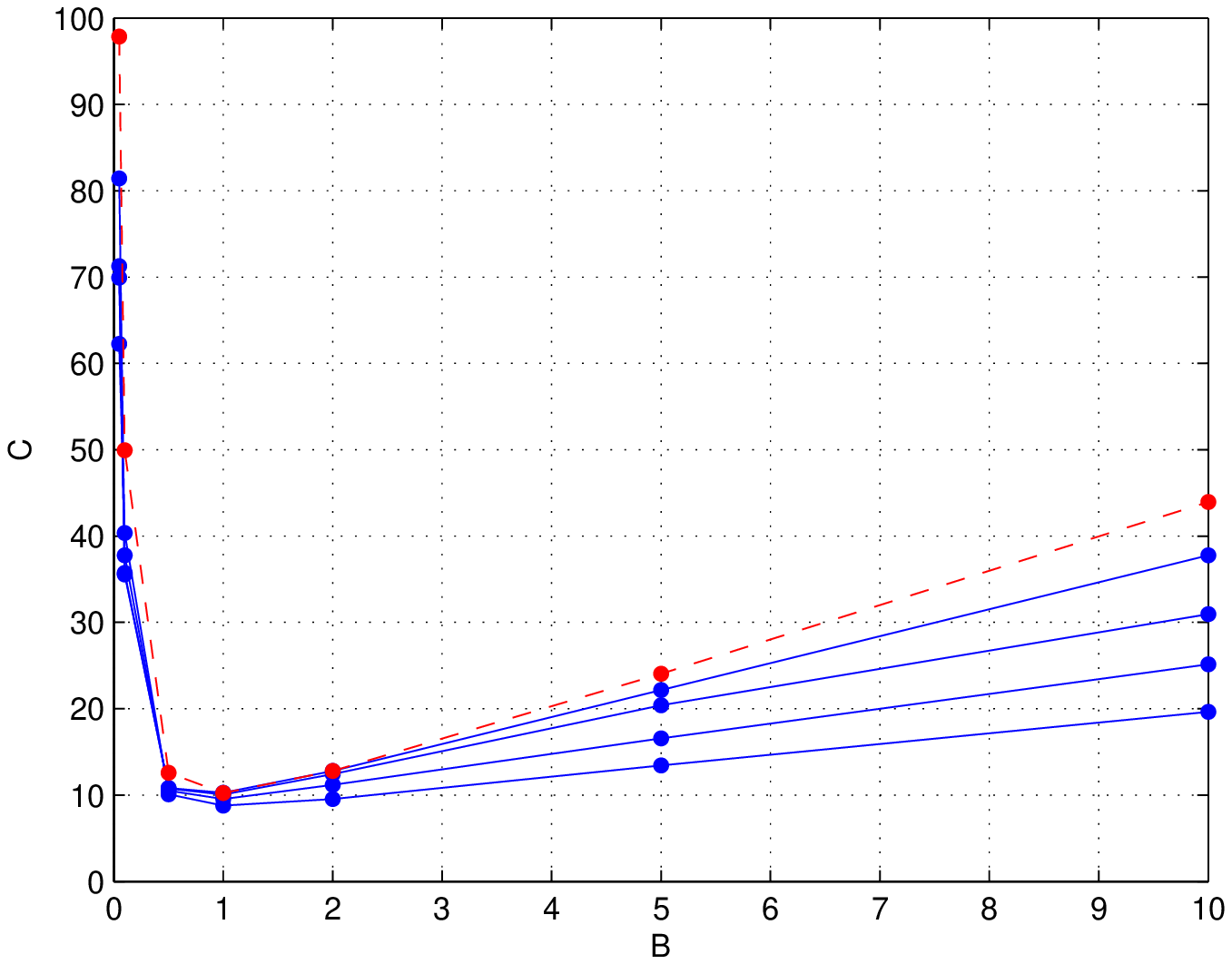}%
  ~\includegraphics[width=0.5\linewidth]{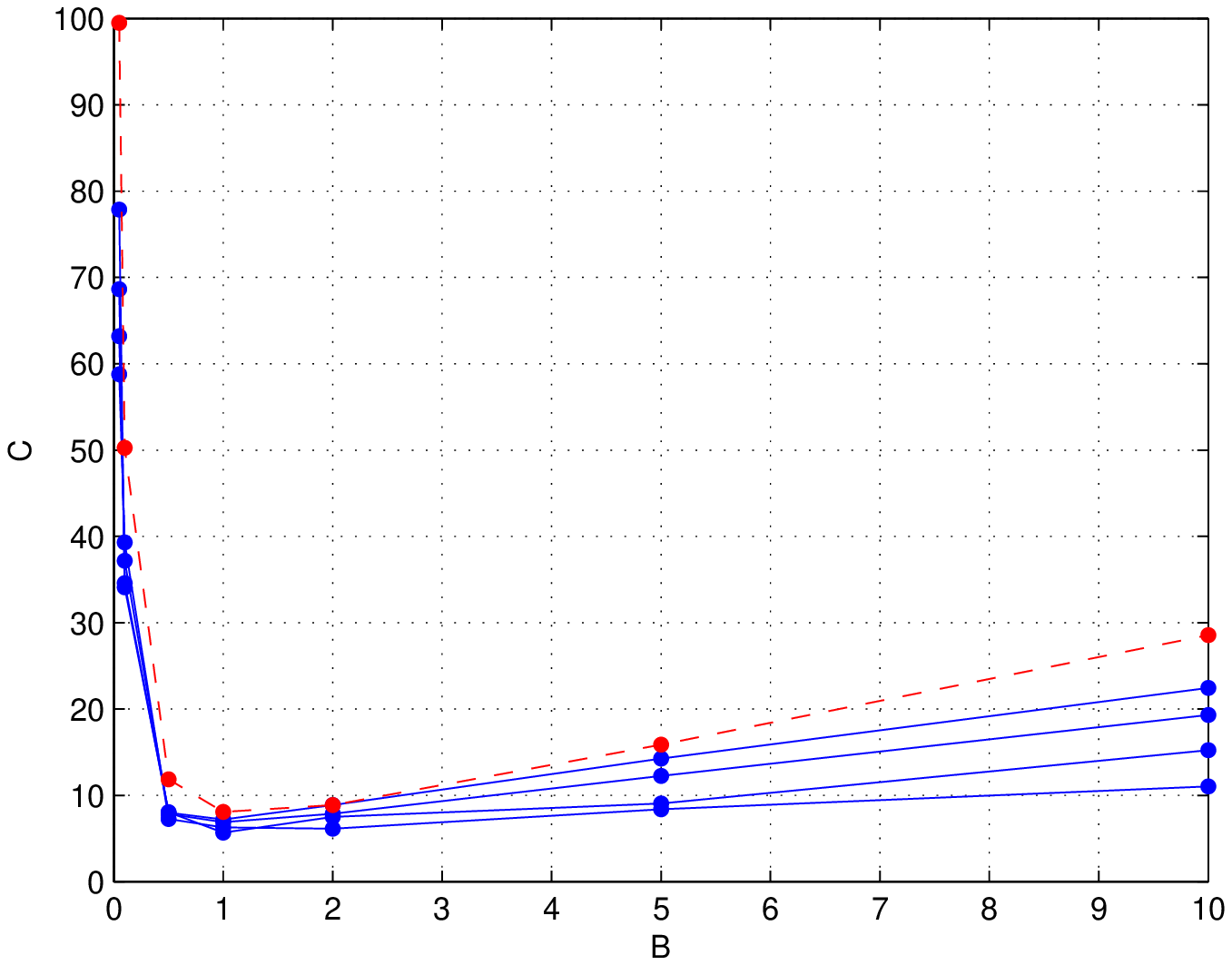}
  \caption{Adiabatic constant as a function of magnetic field strength
    $B$ for different velocities $v=0.01, 0.005, 0.002, 0.001$
    ({\color{blue}\full}, bottom to top) compared to the upper bound
    $\alpha B/\gamma_{\rm min}^2$ ({\color{red}\broken}) for
    anisotropy $\Delta=2$ (left, $\alpha=0.5677$) and $\Delta=10$
    (right, $\alpha=0.6324$) with $L=20$.}
  \label{fig:compare_C_gapfunc}
\end{figure}

We call the smallest constant $C$ in (\ref{equ:adiab}) the adiabatic
constant. Heuristically \cite[17.112]{Mess}, the adiabatic constant is
of the order
\begin{equation}\label{C-heurist}
  \sup_{s\in[0,1]}\frac{\|\frac{d}{ds}{H}(s)\|}{\gamma(s)^2} \,,
  \quad s=t/\tau\,.
\end{equation}
In our case, $H(t)$ is given by (\ref{H(t)}), and
%\begin{equation*}
$\|\frac{d}{ds}{H}(s)\| = \|B S_{x_0}^1 - BS_{x_0+1}^1\| = B$.
%\end{equation*}
Hence, $C \propto B/\gamma_{\mathrm{min}}^2$ with
$\gamma_{\mathrm{min}} = \min_{s\in[0,1]}\gamma(s)$ is the minimal
gap. As $B$ tends to infinity the gap $\gamma$ saturates and therefore
$C$ grows linearly with $B$ for large $B$.  On the other hand, if $B$
is small then the gap shrinks like $B$ (not $B^2$) as has been shown
for a single site perturbation \cite{CNS}, and $C$ diverges as $1/B$.
As a consequence, there is an optimal range for which $C$ is smallest.
In Figure \ref{fig:compare_C_gapfunc} this appears near $B=1$.  If we
fix $B$ and $\D$ and some $\varepsilon$ we can find (empirically) an
upper bound $v_{\max}$ for the velocity $v$ for which then the
adiabatic evolution is $\varepsilon$-close (in the $\ell^2$-sense) to
the true time evolution.  The slowdown (decreased $v_{\max}$) of the
domain wall motion for large $B$-fields was reported
cf.~\cite{beach05}) but here we also observe a slowdown for small
$B$-fields. On a rigorous level, estimates in the vein of
(\ref{C-heurist}) on the adiabatic constant were recently derived by
Jansen, Ruskai, and Seiler~\cite{JRS}.

We have computed the adiabatic constant numerically using adaptive
time-dependent DMRG \cite{vidal2004, white2004}. Unlike the original
method, we express the evolved state in the time-dependent eigenbasis
of the Hamiltonian such that we can compute eq. (\ref{equ:adiab}) with
$\psi(t)$ and $\varphi(t)$ expressed in the same basis (see
\ref{sec:algorithm} for algorithm details).  Figure
\ref{fig:compare_C_gapfunc} shows the adiabatic constant as well as
the heuristic upper bound
\begin{equation*}
  C \leq \alpha \frac{B}{\min_{s\in[0,1]}\gamma(s)^2}\,,
\end{equation*}
with the energy gaps computed by ground state DMRG (see Section
\ref{sec:model}).  To determine $\alpha$, we took one value of $B$
($B=1.0$ for $\Delta=2$ and $B=2.0$ for $\Delta=10$) and made the
inequality in (\ref{C-heurist}) an equality at the lowest $v$
($v=0.001$) at this particular value of $B$.

Since our time-dependent algorithm does not target the state $\psi(t)$
directly, but rather the lowest energy states of the Hamiltonians
$H(t)$ (see \ref{sec:algorithm}), it is important to keep track of how
well $\psi(t)$ is represented in these time-evolving DMRG bases. One
way to measure this is by computing the deviation from $1$ of the norm
of $\psi(t)$. As expected there is a correlation between this norm
loss and the value of the adiabatic constant $C$. In Figure
\ref{fig:compare_C_gapfunc}, where $C$ is minimal, $\|\psi(t)\|$ will
be above $0.999$ at all times; at the very smallest and very largest
$B$-fields, and for the largest velocities still considered adiabatic
($v\approx 0.01)$ the minimum norm drops to about $0.97$.

Adiabaticity is also measured by comparing the value of the total
energy in the time-evolved state with the ground state energy of
$H(t)$, i.e.,
\begin{equation*}
  \frac{\langle \psi(t)| H(t)| \psi(t)\rangle}{\| \psi(t)\|^2} 
  - E_0(t)
\end{equation*}
should be close to $0$ at all times.  For a typical adiabatic speed
$v=0.005$ and intermediate $B$-field, this difference never exceeds
$5\cdot 10^{-4}$. Energy differences in the small and large $B$
regions are of similar magnitude.

Finally, we comment on the role of the interpolation function
(\ref{eq:V}) in our adiabatic Hamiltonian. Let us consider a large
magnetic field.  Then as time proceeds part of the profile is lagging
behind (cfr. Figure \ref{fig:S3_gs}). Also, the change of the profile
is rather quick in the beginning (and at the end) and slow in the
middle of a cycle. To improve on the transport of the domain wall for
a fixed time interval we can take advantage of this phenomenon and
slow down the time evolution in the beginning and accelerate in the
middle by choosing a different interpolation function. In other words,
we may use a general interpolating Hamiltonian $H_f(t) = H_0 - (1-
f(vt)) B S_{x_0}^1 - f(vt) B S_{x_0+1}^1$ with the constraint that
$f(1)=1-f(0) = 1$ to keep the same mean velocity. As an example we
have used the function $f_1(t) = \cos(\pi t/2)$ and a piece-wise
linear function $f_2$ with slope $1/3$ for $t\in[0,1/6]\cup [5/6,1]$
and slope $4/3$ on the interval $[1/6,5/6]$. In the first case we took
$L=20,\D=2,B=5,v=0.05$ and reduced the adiabatic constant from 16.6
(constant velocity, or equivalently linear $f$) to 4.5 calculated with
$f_1$. In the second example with $f_2$ and parameter values
$L=20,\D=2,B=10,v=0.05$ the adiabatic constant dropped from 25.1 to
12.2. This observation is in agreement with \cite{JRS}, where the
adiabatic constant for the general interpolating Hamiltonian $H_f(t)$
was studied.

\section{Fast change of the magnetic field}
\label{sec:fast-change-magnetic}

Now we study the situation when the velocity of the moving magnetic
field is large. %This is based on standard formal results \cite{Mess},
%since we are not aware of rigorous studies in this regime.  
We start initially in the ground state $\phi=\phi_{c_{x_0}}$ of the 
Hamiltonian $H(0)=H_0-BS_{x_0}^1$.
Let $\tau=1/v$ and let $p(\tau)$ be the probability that at time
$\tau$ the system is still in the state $\phi$. Then according to
formula \cite[(17.60)]{Mess},
\begin{equation*}
  p(\tau) = 1 - \tau^2 \,{\rm var}_\phi(\bar{H}) + {\mathcal O}(\tau^3)
\end{equation*}
with ${\rm var}_\phi(\bar{H}) = \langle\phi|\bar{H}^2|\phi\rangle -
\langle\phi|\bar{H}|\phi\rangle^2$, and $\bar{H} = \frac{1}{\tau}
\int_0^\tau H(t)\, dt$. In our example, $\bar{H} = H_0 - \frac{B}{2}
(S_{x_0}^1 + S_{x_0+1}^1)$. Since $\phi$ is also an eigenstate of
$H_0-\frac{B}{2} S_{x_0}^1$ with energy $-\frac{B}4$, we obtain
\begin{eqnarray*}
   \langle\phi|\bar{H}|\phi\rangle =-\frac{B}4 - \frac{B}2
   \langle\phi|S_{x_0+1}^1|\phi\rangle\,,\;
   \langle\phi|\bar{H}^2|\phi\rangle = \frac{B^2}8 +\frac{B^2}4 
   \langle\phi|S_{x_0+1}^1|\phi\rangle \,.
\end{eqnarray*}
Hence,
\begin{equation*}
  {\rm var}_\phi(\bar{H}) = \frac{B^2}{16} \big(1- 
   4\langle\phi|S_{x_0+1}^1|\phi\rangle^2\big)\,.
\end{equation*}
The quantity $\langle\phi|S_{x_0+1}^1|\phi\rangle$ can be computed fairly
explicitly as a function of $\D$ but we are content with the trivial estimate
that $1 - 4\langle\phi|S_{x_0+1}^1|\phi\rangle^2 \le 1$. As a result, the 
probability to stay in the initial state $\phi$ until $\tau$,
\begin{equation*}
   p(\tau) \ge 1-\frac{\tau^2 B^2}{16} + {\mathcal O}(\tau^3) \,.
\end{equation*}
If we want this to be larger than $1-\varepsilon$, then $|B|$ needs to be
smaller than $4v\sqrt{\varepsilon} $. 

\begin{figure}
  \centering
  \includegraphics[width=0.5\linewidth]{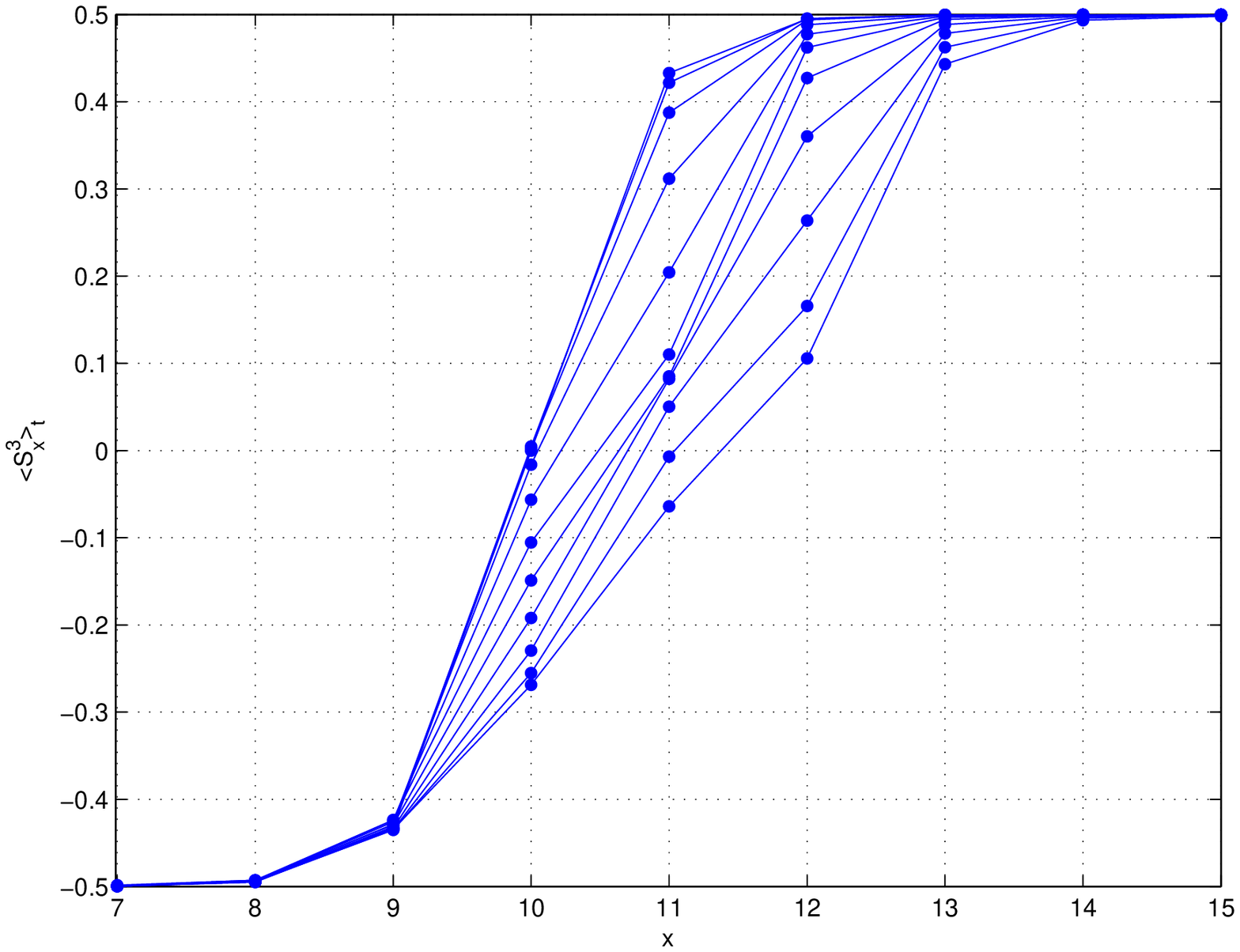}%
  ~ \includegraphics[width=0.5\linewidth]{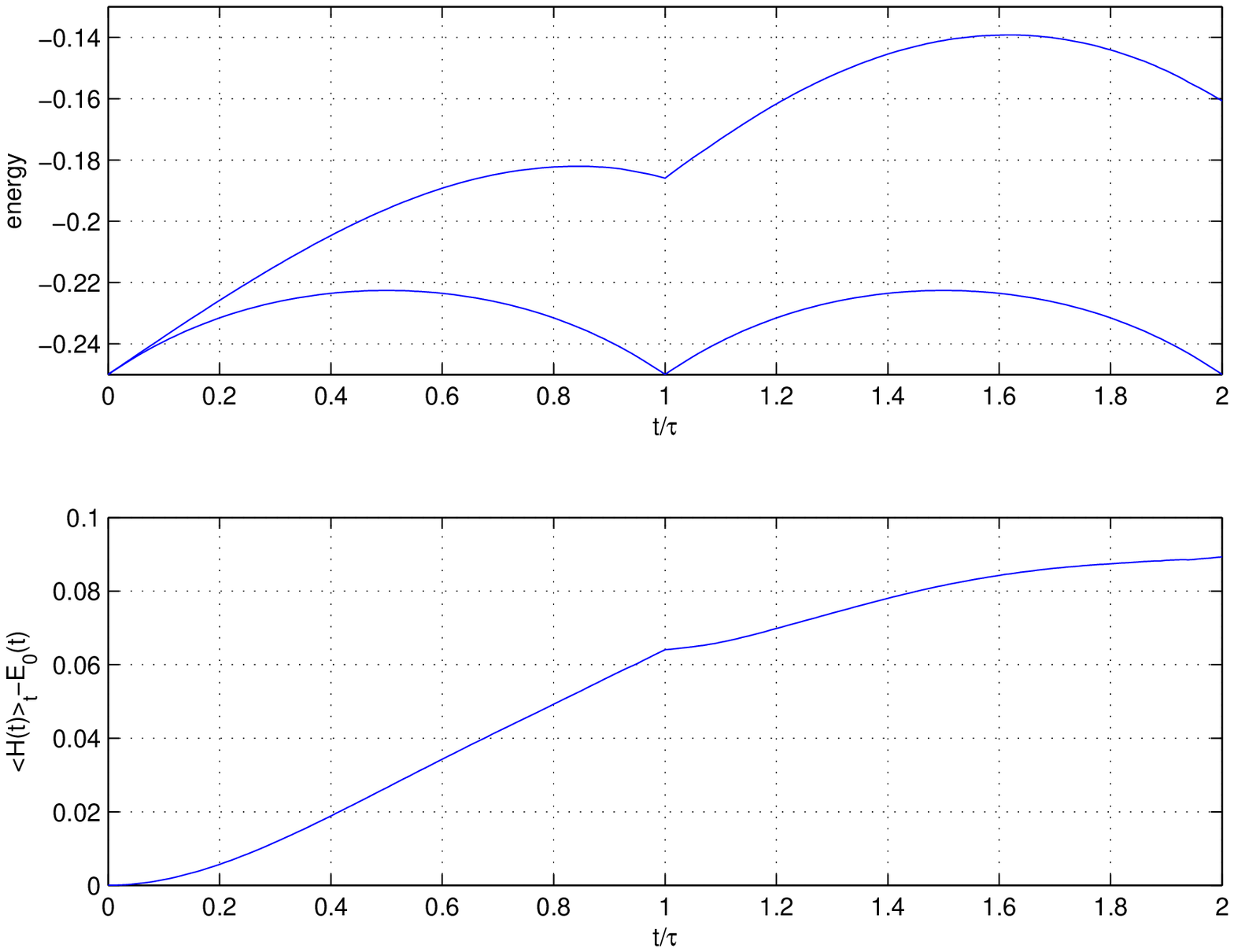}
  \caption{$S^3$-magnetization profiles of $\psi(t)$, for $0\leq t\leq
    \tau=2v^{-1}$ (left) and energy of the ground state and of
    $\psi(t)$ (right, top) and energy difference between $\psi(t)$ and
    the ground state (right, bottom) ($\Delta=2$, $B=0.5$, $v=0.1$,
    $200$ time steps, $S^3$-profiles plotted for every $20$ time
    steps).}
  \label{fig:S3-profile-non-ad}
\end{figure}

Numerically we can also investigate the intermediate region between
adiabatic and sudden change of the magnetic field. Here, $B$ and $v$
are of the same order of magnitude.  In Figure
\ref{fig:S3-profile-non-ad}, we follow the $S^3$-magnetization profile
of the time evolved state over two periods, with the natural extension
of $V(t)$ in (\ref{eq:V}) such that the field keeps moving to the
right.  Although the position of the interface is transported in the
same direction as the magnetic field, it is lagging behind w.r.t.  the
postion of the ground state interfaces $\varphi(t)$ (which moves from
site $10$ to site $12$). At the same time the width of the interface
is growing bigger. Also the energy of $\psi(t)$ is lagging behind
w.r.t.  the periodicity of the spectrum of $H(t)$ (see Figure
\ref{fig:S3-profile-non-ad}). The difference with the ground state
energy is steadily increasing, and it is an interesting question
whether this difference will eventually saturate.  In this
computation, the minimum norm of $\psi(t)$ remains above $0.99$ for
the first period ($t\leq v^{-1}$) and above $0.97$ for the second
period ($t\leq 2v^{-1}$)), so despite the non-adiabatic transport,
$\psi(t)$ is still well represented in the DMRG basis constructed from
the low-energy spectrum of $H(t)$. The overlap with the ground state
$\varphi(t)$ drops to $0.80$ during the first period, and further to
$0.71$ during the second period.

We have tested other magnetic field perturbations such as a smooth
field
\begin{equation*}
  V(t) = \frac{B}4\sum_x \bigl(1+\cos(\pi(x-vt))\bigr)^2\chi_L
  (x-vt)\, S^1_x
\end{equation*}
where
\begin{equation*}
  \chi_L(x) = \cases{
    1 &  for  $x \in [L/2-1,L/2+1]$\\
    0 &  otherwise 
  }
\end{equation*}
For $L$ a multiple of $4$, this is again a single-site perturbation
for $t=nv^{-1}$, $n=0,1,2,\dots$, and interpolating smoothly in
between. In this case, the non-differentiable curves in Figure
\ref{fig:S3-profile-non-ad} become smooth, but no other qualitative
differences are observed.

Let us summarize the two regimes in terms of the two parameters $v$
and $B$.  We are in the adiabatic regime if $C v\ll1$. Since for small
$B$, the adiabatic constant $C$ is inverse proportional to $B$, we
require $v\ll B$. For large $B$ we know that $C$ is proportional to
$B$ and thus we want that $v\ll 1/B$.  The regime where the initial
state is stationary is simply given by the condition that $B\ll v$.

\section{Summary} 

We have studied the propagation of a magnetic domain wall in the 
ferromagnetic Heisenberg chain due to a time dependent magnetic field. 
This situation is of immediate interest in submicrometer magnetic 
wires~\cite{beach05,Ono} and in establishing ferromagnetic 
gates~\cite{Allwood}. In the simplest case, the magnetic field is 
localized near the center of the domain wall (say at site $x_0$) and 
then moved along the chain to site $x_1$. If its velocity is small 
then we are in the adiabatic regime and the domain wall is shifted 
from $x_0$ to $x_1$ thereby preserving its shape. We find that there
is an optimal region of the strength of $B$-fields for which the domain
wall mobility is highest and is reduced for large and small $B$-fields.

On top of an analytical study using the standard Adiabatic 
Theorem and based upon properties of the Heisenberg model that have been 
proved in recent years, we have performed a numerical study. The latter 
is a variation of the recently developed time-dependent Density Matrix 
Renormalization Group method. This allows us to follow the Schr\"odinger 
time evolution of the domain wall with high accuracy in the adiabatic as 
well as the non-adiabatic regime, and discuss in detail the dependence 
on the parameters of our model such as the strength and speed of the 
magnetic field and the coupling constant in the Heisenberg interaction.

\appendix

\section{DMRG algorithms}
\label{sec:algorithm}

The DMRG algorithms for ground state \cite{white1993} as well as
time-dependent \cite{vidal2004,Daley,white2004,gobert05} computations
are well known, see also \cite{schollwoeck2005} and \cite{peschel99}.
However, degeneracy and non-translation invariance of the ground
states of the (translation invariant) XXZ kink Hamiltonian require a
few non-standard adaptations. To lift degeneracy, we restrict to a
sector of fixed total $S^3$-magnetization. For a chain of even length,
the ground state with total magnetization zero has an interface
centered in between the two middle sites. This state can be grown in
the usual way by successively inserting two sites in the middle and
targeting the zero magnetization state at each step. For a kink state
with non-zero magnetization, we grow the system initially in the zero
magnetization sector. If eventually a total magnetization with
interface close to the middle is needed, we perform the finite system
convergence sweeps in the new sector, with the zero magnetization
state as initial trial state. If an interface close to one of the
edges is needed, we grow the system at the right moment by targeting a
magnetization sector which increases or decreases with one each step
(effectively inserting two up or two down spins in the middle, thus
shifting the kink to the right or left).  The DMRG enlarging process
implicitly assumes a translation invariant Hamiltonian. However, if
the breaking of translation invariance is due to a local perturbation
as in our case (with external magnetic fields located on one or two
sites), we can grow the system targeting the zero magnetization kink,
and add the perturbation for the finite system convergence sweeps
where translation invariance is no longer needed.  The DMRG state thus
obtained converges indeed very rapidly to the true ground state
$\psi(0)$ of the perturbed Hamiltonian \eref{H(t)} at time $t=0$.

In the standard time-dependent DMRG, adaptation of the basis is done
by computing and truncating new reduced density matrices for the
evolved state $\psi(t)$. Here the situation is different because we
need to compute both the time evolved state $\psi(t)$ and the true
ground state $\varphi(t)$ of $H(t)$ (in the same DMRG basis,
naturally). Therefore we use the ground state (or several low-lying
states) of $H(t)$ to adapt the Hilbert spaces.  The details of our
algorithm are as follows.  The time period $[0,\tau]$ is first divided
in $n_{\tau}$ time steps of length $\delta=\tau/n_{\tau}$ by
discretizing the time evolution, i.e., the evolved states $\psi(t)$
are defined by
\begin{eqnarray*}
  \psi(0) &= \varphi(0)\,,\\
  \psi(n\delta) &= \e^{-\i\delta H(n\delta)}\psi((n-1)\delta)\,,\;
  n=1,\dots,n_{\tau}\,.
\end{eqnarray*}
Each of these time steps is further divided into $n_T$ smaller steps
of length $\delta_T=\delta/n_T$ for the Trotter decomposition, so
\begin{eqnarray*}
  \psi(n\delta) &\approx \bigl(\e^{-\i\delta_T H(n\delta)}\bigr)^{n_T}
  \psi((n-1)\delta)\,,
\end{eqnarray*}
and $\e^{-\i\delta_T H(n\delta)}$ is expanded by
\begin{equation*}
  \e^{-\i\delta_T H} \approx \e^{-\frac{\i}2\delta_T h_{1,2}}
  \e^{-\frac{\i}2\delta_T h_{2,3}} \cdots \e^{-\i\delta_T h_{L-1,L}}
  \cdots \e^{-\frac{\i}2\delta_T h_{2,3}} \e^{-\frac{\i}2\delta_T h_{1,2}}\,,
\end{equation*}
such that each factor acts on two sites which are successively
represented exactly in a DMRG sweeping process.  Note that only the
interaction $h_{L/2,L/2+1}$ is time-dependent. We shall call the so
approximated state $\psi_T(n\delta)$.  Let us start at time $0$ where
$\psi(0) = \varphi(0)$. To apply $\e^{-\i\delta H(\delta)}$ to it we
need to have $\psi(0)$ written in a DMRG basis that represents the low
energy states of $H(\delta)$. To this end we use $\varphi(0)$ as an
initial trial state and apply standard finite system DMRG sweeps
targeting the ground state $\varphi(\delta)$ of $H(\delta)$, and
update $\psi(0)$ along the way. At the end of the sweeps we have a new
DMRG basis for $H(\delta)$, its ground state $\varphi(\delta)$, and
$\psi(0)$ expressed in the new basis. This is the \emph{adaptive} part
of our algorithm.  Next we can apply $\e^{-\i\delta H(\delta)}$ to the
new representation of $\psi(0)$ using the Trotter sweeping process and
obtain the evolved state $\psi_T(\delta)$. By construction
$\psi_T(\delta)$ and $\varphi(\delta)$ are expressed in the same DMRG
basis, and their overlap and other quantities can be computed.  The
subsequent time steps proceed in exactly the same manner.

Although the motivation for adapting the DMRG bases in this way is
inspired by the question to analyze the adiabatic approximation, we
have found that it is also very convenient to compute the time-evolved
state in the non-adiabatic regime. In this case, we need to target
more low-energy states, but we do not need to increase the block
dimension, unlike standard time-dependent DMRG which needs
approximately twice the block dimension of ground state DMRG to
achieve the same accuracy \cite{white2004}.

\section{Parameter values and software}

Our problem is suitable for DMRG with small block dimension as its
entanglement entropy \cite{vidal2003} goes to zero for large intervals
around the support of the magnetic field perturbation (it is exactly
zero for the non-perturbed Hamiltonian due to its frustration free
property). The algorithm converges up to machine precision to the
true, known energies of the ground state and first excited state with
the number of kept states (block dimension) as low as $16$ for a chain
of length $L=20$. We carried out 3 system sweeps to obtain convergence
of the ground states and of the DMRG bases at each time step. For
given $v$, the final time $\tau=1/v$ was divided in time steps of
length $1$ which were further subdivided into $n_T=100$ Trotter steps.
For the ground state and adiabatic transport computations, we targeted
the $3$ lowest energy states of $H(t)$ to construct the DMRG basis,
increasing to the $5$ lowest energy states for the fast changing
magnetic field. A straightforward error analysis shows that $\big| C_T
-C\big| = {\mathcal O}(\frac{\xi}{v})$, where $C_T$ it the adiabatic
constant computed using the Trotter approximation to the time-evolved
state, $C$ is the true adiabatic constant, and $\xi= 1/(v
n_T^2)\max(1,B^2)$.

We have implemented the ground state and time-dependent DMRG
algorithms in a Matlab Toolbox which can be easily applied to other
models as well.  The software is included in the tar archive with the
\LaTeX{} source and figure files of the paper, available for download
at \texttt{arXiv:cond-mat/0702059}.

\ack T.M. is a Postdoctoral Fellow of the Research Foundation --
Flanders (Belgium). The research of B.N. is supported in part by U.S.
National Science Foundation grant \# DMS-06-05342.  W.S. is grateful
for the support and hospitality at the International University
Bremen, Germany.

\section*{References}

%\bibliographystyle{unsrt}
%\bibliography{xxz}

\begin{thebibliography}{10}

\bibitem{Allwood}
D.A Allwood, Gang Xiong, M.D. Cook, C.C. Faulkner, D.~Atkinson, N.~Vernier, and
  R.P. Cowburn.
\newblock Submicrometer ferromagnetic not gate and shift register.
\newblock {\em Science}, 296:2003--2006, 2002.

\bibitem{beach05}
G.S.D. Beach, C.~Nistor, C.~Knutson, M.~Tsoi, and J.L. Erskine.
\newblock Dynamics of field-driven domain-wall propagation in ferromagnetic
  nanowires.
\newblock {\em Nature}, 4:741--744, 2005.

\bibitem{Ono}
T.~Ono, H.~Miyajima, K.~Shigeto, K.~Mibu, N.~Hosoito, and T.~Shinjo.
\newblock Propagation of a magnetic domain wall in a submicrometer magnetic
  wire.
\newblock {\em Science}, 284:468--470, 1999.

\bibitem{hirjibehedin06}
C.F. Hirjibehedin, C.P. Lutz, and A.J. Heinrich.
\newblock Spin coupling in engineered atomic structures.
\newblock {\em Science}, 312:1021--1024, 2006.

\bibitem{white1993}
S.~R. White.
\newblock Density-matrix algorithms for quantum renormalization groups.
\newblock {\em Phys. Rev. B}, 48(14):10345--10356, 1993.

\bibitem{peschel99}
I.~Peschel, X.~Wang, M.~Kaulke, and K.~Hallberg, editors.
\newblock {\em Density-Matrix Renormalization - A New Numerical Method in
  Physics}, volume 528 of {\em Lecture Notes in Physics}.
\newblock Springer, 1999.

\bibitem{vidal2004}
G.~Vidal.
\newblock Efficient simulation of one-dimensional quantum many-body systems.
\newblock {\em Phys. Rev. Lett.}, 93(4):040502, 2004.

\bibitem{Daley}
A.~J. Daley, C.~Kollath, U.~Schollw\"ock, and G.~Vidal.
\newblock Time-dependent density-matrix renormalization-group using adaptive
  effective {H}ilbert spaces.
\newblock {\em J. Stat. Mech.: Theor. Exp.}, page P04005, 2004.

\bibitem{white2004}
S.~R. White and A.~E. Feiguin.
\newblock Real time evolution using the density matrix renormalization group.
\newblock {\em Phys. Rev. Lett.}, 93:076401, 2004.

\bibitem{gobert05}
D.~Gobert, C.~Kollath, U.~Schollw\"ock, and G.~Sch\"utz.
\newblock Real-time dynamics in spin-(1/2) chains with adaptive time-dependent
  density matrix renormalization group.
\newblock {\em Phys. Rev. E}, 71:036102, 2005.

\bibitem{FNW1}
M.~Fannes, B.~Nachtergaele, and R.~F. Werner.
\newblock Finitely correlated states of quantum spin chains.
\newblock {\em Commun. Math. Phys.}, 144:443--490, 1992.

\bibitem{PS}
V.~Pasquier and H.~Saleur.
\newblock Common structures between finite systems and conformal field theories
  through quantum groups.
\newblock {\em Nuclear Physics B}, 330:523--556, 1990.

\bibitem{ASW}
F.~C. Alcaraz, S.~R. Salinas, and W.~F. Wreszinski.
\newblock Anisotropic ferromagnetic quantum domains.
\newblock {\em Phys. Rev. Lett}, 75:930--933, 1995.

\bibitem{GW}
C.-T. Gottstein and R.~F. Werner.
\newblock Ground states of the infinite q-deformed {H}eisenberg ferromagnet.
\newblock {\em Eprint arXiv:cond-mat/9501123}, 1995.

\bibitem{CNS}
P.~Contucci, B.~Nachtergaele, and W.L. Spitzer.
\newblock The ferromagnetic {H}eisenberg {XXZ} chain in a pinning field.
\newblock {\em Phys. Rev. B}, 66:0644291--13, 2002.

\bibitem{Gwa1992}
L.~H. Gwa and H.~Spohn.
\newblock Six-vertex model, roughened surfaces, and an asymmetric spin
  hamiltonian.
\newblock {\em Phys. Rev. Lett.}, 68:725--728, 1992.

\bibitem{Derrida1993}
B.~Derrida, M.R. Evans, V.~Hakim, and V.~Pasquier.
\newblock Exact solution of a 1d asymmetric exclusion model using a matrix
  formulation.
\newblock {\em J. Phys. A}, 26:1493--1517, 1993.

\bibitem{Sandow1994}
S.~Sandow and G.~Sch\"utz.
\newblock On $u_q[su(2)]$-symmetric driven diffusion.
\newblock {\em Europhys. Lett.}, 26:7--12, 1994.

\bibitem{KN1}
T.~Koma and B.~Nachtergaele.
\newblock The spectral gap of the ferromagnetic {XXZ} chain.
\newblock {\em Lett. Math. Phys.}, 40:1--16, 1997.

\bibitem{KN3}
T.~Koma and B.~Nachtergaele.
\newblock The complete set of ground states of the ferromagnetic {XXZ} chains.
\newblock {\em Adv. Theor. Math. Phys.}, 2:533--558, 1998.

\bibitem{NS}
B.~Nachtergaele and S.~Starr.
\newblock Droplet states in the {XXZ} {Heisenberg} model.
\newblock {\em Commun. Math. Phys.}, 218:569--607, 2001.

\bibitem{NSS3}
B.~Nachtergaele, W.~Spitzer, and S.~Starr.
\newblock Ferromagnetic ordering of energy levels.
\newblock {\em Journ. Stat. Phys.}, 116:719--738, 2004.

\bibitem{NSS4}
B.~Nachtergaele, W.~Spitzer, and S.~Starr.
\newblock {Droplet Excitations for the Spin-$1/2$ XXZ Chain with Kink oundary
  Conditions}.
\newblock To appear in Ann. Henri Poincar\'e; DOI {10.1007/s00023-006-0304-6},
  2006.

\bibitem{Thi3}
W.~Thirring.
\newblock {\em Quantum Mathematical Physics: Atoms, Molecules and Large
  Systems}.
\newblock Springer, 2003.

\bibitem{Mulherkar2007}
J.~Mulherkar, B.~Nachtergaele, R.~Sims, and S.~Starr.
\newblock Isolated eigenvalues of the ferromagnetic spin-{$J$} {XXZ} chain with
  kink boundary conditions.
\newblock {\em Eprint arXiv:0709.1733}, 2007.

\bibitem{Kitaev}
A.Yu. Kitaev, A.H. Shen, and M.N. Vyalyi.
\newblock {\em Classical and Quantum Computation}, volume~47 of {\em Graduate
  Studies in Mathematics}.
\newblock American Mathematical Society, 2002.

\bibitem{Kato}
T.~Kato.
\newblock On the adiabatic theorem of quantum mechanics.
\newblock {\em Journ. Phys. Soc. Jap.}, 5:435--439, 1955.

\bibitem{Mess}
A.~Messiah.
\newblock {\em Quantum Mechanics}.
\newblock Dover Publications, 1999.

\bibitem{JRS}
S.~Jansen, M.B. Ruskai, and R.~Seiler.
\newblock Bounds for the adiabatic approximation with applications to quantum
  computation.
\newblock preprint.

\bibitem{schollwoeck2005}
U.~Schollw\"ock.
\newblock The density-matrix renormalization group.
\newblock {\em Rev. Mod. Phys.}, 77:259, 2005.

\bibitem{vidal2003}
G.~Vidal, J.I. Latorre, E.~Rico, and A.~Kitaev.
\newblock Entanglement in quantum critical phenomena.
\newblock {\em Phys. Rev. Lett.}, 90(22):227902, 2003.

\end{thebibliography}

\end{document}